# Lateral movements of individual AMPA receptors inside synapses contribute to the regulation of receptor accumulation.

Catherine Tardin*†°, Laurent Cognet*°, Cécile Bats†, Brahim Lounis* & Daniel Choquet†.

*Centre de Physique Moléculaire Optique et Hertzienne - CNRS UMR 5798 et Université Bordeaux 1, 351 Cours de la Libération, 33405 Talence, France*

*† Laboratoire de Physiologie Cellulaire de la Synapse - CNRS UMR 5091 et Université Bordeaux 2, Institut François Magendie, 1 rue Camille Saint-Saëns 33077 Bordeaux, France*

*° These authors contributed equally to this work*

Correspondence and requests for materials should be addressed to D.C. (e-mail: dchoquet@u-bordeaux2.fr).

Total character count : 57477

**Trafficking of AMPA receptors in and out of synapses is crucial for synaptic plasticity. Previous studies have focused on the role of endo/exocytosis processes or that of lateral diffusion of extrasynaptic receptors. We now directly imaged AMPAR movements inside and outside synapses of live neurons using single-molecule fluorescence microscopy. Inside individual synapses, we found immobile and mobile receptors, which display restricted diffusion. Extrasynaptic receptors display free diffusion. Receptors could also exchange between these membrane compartments through lateral diffusion. Glutamate application increased both receptor mobility inside synapses and the fraction of mobile receptors present in a juxtasynaptic region. Block of inhibitory transmission to favor excitatory synaptic activity induced a transient increase in the fraction of mobile receptors and a**

**decrease in the proportion of juxtasynaptic receptors. Altogether, our data show that rapid exchange of receptors between a synaptic and extra-synaptic localization occurs through regulation of receptor diffusion inside synapses.**

**INTRODUCTION**

AMPA glutamate receptors (AMPARs) are ligand-activated cation channels concentrated in the postsynaptic density (PSD) which mediate fast excitatory neurotransmission in the central nervous system (Dingledine et al., 1999). Concentration of AMPARs at PSDs is thought to result from their stabilization by interactions with specific intracellular scaffolding proteins and cytoskeletal elements (Nusser, 2000; Scannevin and Huganir, 2000). In fact, AMPARs constitutively cycle in and out of the postsynaptic membrane at a rapid rate through processes of endo- and exocytosis (Carroll et al., 2001; Ehlers, 2000; Luscher et al., 1999; Noel et al., 1999). The turnover rate of the whole cycle is regulated by neuronal activity and scaffolding or accessory proteins such as GRIP and NSF (reviewed in (Barry and Ziff, 2002; Braithwaite et al., 2000; Carroll et al., 2001; Malinow and Malenka, 2002; Sheng and Kim, 2002)). Regulation of the balance between endo- and exocytosis of AMPARs may account for the rapid variations in receptor composition of the PSD during synaptic plasticity. A simplified scheme would be that postsynaptic LTP involves increased exocytosis of AMPARs while LTD would be underlied by increased endocytosis of receptors.

Recently, several pieces of evidence have also indicated a role of receptor diffusion within the plasma membrane in AMPARs trafficking in and out of PSDs. First, GluR2 containing AMPARs diffuse rapidly in the extrasynaptic membrane and stop reversibly in the periphery of synapses (Borgdorff and Choquet, 2002). Second, some processes of endo- or exocytosis occur at the periphery of synapses rather than directly at PSDs. For endocytosis, clathrin assembly and disassembly occurs at hotspots throughout dendrites laterally to the PSDs (Blanpied et al., 2002). LTD or glutamate application trigger an increase in the amount of endocytosed AMPARs (Beattie et al., 2000; Carroll et al., 1999a; Carroll et al., 1999b; Wang and Linden, 2000). However, glutamate by itself does not increase the efficiency of the endocytotic pathway (Zhou et al., 2001). Further, depolymerization of the actin cytoskeleton causes internalization of AMPARs, whereas a drug that stabilizes actin filaments blocks internalization. Glutamate could thus induce a dissociation of AMPARs from their anchors in the PSD and this may be followed by entry of AMPARs in a constitutive endocytotic pathway in the extra synaptic domain of the membrane. Altogether, this indirectly points out to the hypothesis that endocytosis of AMPARs may first require their dispersal from the synaptic to the extra synaptic membrane through lateral diffusion.

Reciprocally, whether there are also hotspots for exocytosis of receptors outside of the PSD is unknown. Using cleavable extracellular tags, it was established that at early times after exocytosis, new GluR1 subunits are diffusively distributed along dendrites. This is followed by their lateral translocation and accumulation into synapses (Passafaro et al., 2001). Delivery of receptors to the PSD after exocytosis may thus also involve diffusion. Interestingly, GluR2 accumulated faster than GluR1 at synaptic sites.

Mechanisms involved in the regulation of the accumulation of AMPARs at synaptic sites through lateral diffusion are emerging. Local increases in intracellular calcium drastically reduce AMPARs diffusion rate. At the molecular level, stargazin, a protein which links AMPARs to PSD-95, might regulate AMPARs trafficking between the synaptic and extra synaptic membrane. Stargazin could help to trap diffusing AMPARs by binding to synaptic PSD-95 (Chen et al., 2000; Schnell et al., 2002). Therefore, it is likely that diffusion of receptors in the plane of the membrane is necessary for their removal or addition to and from the PSD, although this has never been directly visualized.

If receptors enter and leave synapses through lateral diffusion, they have to unbind from the postsynaptic scaffold, diffuse through the PSD and exit the synapse. Clusters of receptors somewhat mimicking the PSD can be induced in the synaptic and extra-synaptic membrane by co-expression of scaffold and receptor molecules. At the extra-synaptic clusters, we had directly visualized entry and exit of receptors from the clusters at high rates using single particle tracking (Meier et al., 2001; Sergé et al., 2002). However, the size of the particle (500 nm) precluded tracking of receptors inside the synaptic cleft.

In the present work, we use the single-molecule fluorescence imaging approach (Dickson et al., 1996; Schutz et al., 2000; Seisenberger et al., 2001; Ueda et al., 2001) and review in (Special-Issue, 1999) to localize and track GluR2-containing AMPA receptors inside synaptic sites below the optical diffraction limit. We show that a large proportion of AMPARs diffuse inside synapses and that this diffusion is regulated during protocols that modify receptor accumulation at synapses. We propose that receptor disappearance from postsynaptic sites involve their dispersal through increased lateral diffusion while receptor accumulation involves their delayed stabilization after diffusion.

## RESULTS
### Single molecule imaging in live neurons

Anti-GluR2 antibodies were labeled with Cy5 or Alexa-647 molecules at low labeling ratio (mean labeling ratio of 0.4 dye per antibody) so that individual antibodies were labeled at most with one fluorophore. A small proportion of surface expressed AMPA receptors containing the GluR2 subunit were selectively labeled in live neurons through short incubations with these antibodies. We could thus image and resolve discrete fluorescence spots with an epifluorescence imaging setup (Dickson et al., 1996; Schmidt et al., 1995). The majority of the fluorescence spots (75 ± 6 %, n=80 neurons) exhibit one-step photobleaching (Fig. 1E, movie #1-4) and not a gradual decay as for ensemble photobleaching. The width of these spots corresponds to the point-spread function of the microscope and the signal originating from them ranges from 500 to 1000 counts per 30 ms (Schutz et al., 2000). Thus, these fluorescence spots have all the hallmarks of individual fluorescent molecules (see Fig1S in supplementary materials and reviews in (Special-Issue, 1999)) bound to GluR2 receptors. Only these spots were thus retained for analysis. The imaged single molecules were optically well resolved (Fig. 1C and movie #1-4) and their density on the cell surface was much less than 1 $\mu m^{-2}$. This indicates that antibody incubation did not result in cross-linking of more than two GluR2-containing AMPARs, the anti-GluR2 being bivalent. This was further supported by immunocytochemistry experiments: the apparent level of receptor clustering was smaller when incubation with anti-GluR2 was performed on live when compared to fixed cells (percentage of clustered receptors 15 % ± 7, n=12 and 23% ± 9, n=12, respectively). However this does not rule out that single molecule tracking follows the movement of a natural cluster of receptors, only one receptor being labeled.

Trajectories of such molecules were reconstructed from image series recorded at a rate of 33 Hz (see movies in supplementary data). The length of the trajectories varied from 0.1- 0.5 seconds up to 4 seconds, depending on the photobleaching time of the molecule (mean ± S.D. 244 ± 318 ms, n=3078 molecules). The mean-square-displacement corresponding to trajectories of individual fluorescent molecules dried on glass shows that individual molecules are pointed within 45±5 nm accuracy (Schmidt et al., 1995; Thompson et al., 2002) (Fig. 2B, trajectory #1).

**GluR2 molecules are imaged in synapses**

We first analyzed the spatial distribution of AMPARs with respect to synaptic sites in bulk immunocytochemistry experiments and at the single molecule level in live neurons. For both types of experiments, live neurons were incubated for short periods with anti-GluR2 antibodies (10 minutes). Only for bulk visualization of receptors, this step was followed by fixation and amplification of the signal through secondary antibodies. In immunocytochemistry experiments, AMPARs accumulated in front of glutamatergic presynaptic terminals specifically stained by the vesicular glutamatergic transporter BNPI/VGLUT1 (Fig. 1F). A similar accumulation was previously observed using other presynaptic markers (Carroll et al., 1999b; Noel et al., 1999; Snyder et al., 2001). Single molecule imaging allowed us to further localize AMPARs in live neurons below the optical diffraction limit. Presynaptic terminals were stained with FM1-43 or rhodamine 123 (Fig. 1B and movie #2-4). We measured the distance $r$ between each individual AMPARs and the center of the closest stained synaptic site. We plotted $S(r)$, the proportion of individual molecules per unit surface, as a function of $r$ (Fig. 1G). Individual AMPARs are strongly enriched (about 10 times) at and close to (<300-400nm) synaptic sites. On average, the concentration of AMPARs appears lower at the center of synapses, as sometimes observed at the electron microscopy level (Takumi et al., 1999), although this is not statistically significant. Altogether, these experiments indicate that synaptic AMPARs can be stained through incubation with antibodies in live neurons. They further establish that live staining of presynaptic terminals is a valid approach to discriminate between synaptic and extra-synaptic regions and analyze AMPARs movements in these membrane domains.

**Diffusion characteristics are correlated with localization with respect to synapses: Examples**

Single molecule trajectories were initially sorted into two main categories: the first one corresponds to molecules located at a distance smaller than 300 nm from the center of a presynaptic staining and are referred to as "synaptic" through all this work using this cut-off criterion (see methods). Three of these trajectories are illustrated by examples #2 and #3 in Fig 2A and in movie #4. The second group category corresponds to all the others, including molecules found in the periphery of synapses. They are referred to as "extra-synaptic" and illustrated by examples #4 in Fig 2A and movies. We observed a variety of mobility behaviors for the receptors, which are correlated with the localization. They ranged from highly mobile receptors only seen in extra-synaptic regions (exemplified by trajectory #4 on Fig. 2A) to mildly mobile (trajectory #3 on Fig. 2A) or immobile (trajectory #2 on Fig. 2A). The two latter were mainly found at synaptic sites. Strikingly, individual receptors directly entering and leaving synaptic domains could be observed on occasions ( ~ 1-2 % of all trajectories, e.g. trajectory #5 on Fig. 2A and movie #2).

The mean-square-displacement (MSD) is widely used to extract diffusion characteristics from the trajectories. In Fig. 2B, the MSD is plotted as a function of time for trajectories #1 to #4 illustrated in Fig. 2A. The extra-synaptic mobile receptor diffused freely, as indicated by its linear MSD, while the movement of mobile receptors in synapses was confined to a domain, as indicated by the plateau reached by its MSD over time. The domain size was typical for a synapse (~ 400 nm, see below).

**Diffusion characteristics are correlated with localization with respect to synapses: statistical and analytical analysis**

We performed a statistical analysis on the mobility of AMPARs in each region (synaptic or extra-synaptic). We first analyzed the trajectories of AMPARs in the presence of TTX to block spontaneous neuronal activity (the examples shown on Fig 2A and 2b were part of 493 trajectories from more than 20 different neurons recorded at 37°C). After having sorted the 493 trajectories with respect to their localizations in the two regions, we calculated the instantaneous diffusion coefficient, D, for each trajectory, from linear fits of the first 3-5 points (corresponding to 90 to 150 ms) of the MSD (Anderson et al., 1992) using MSD $(\tau)=<r^2>(\tau)=4D\tau$.

Distributions of D for synaptic and extra-synaptic receptors were strikingly different (Fig. 2C and 2D). Overall, synaptic receptors diffused more slowly than extra-synaptic ones. In the extra-synaptic region, we distinguished three populations from the histograms (Fig. 2C). For the majority (66%) of the molecules, the diffusion coefficient was in the range of 0.1-1 $\mu m^2/s$ (0.45±0.05 $\mu m^2/s$, n=202 trajectories, 7 experiments; data are given as mean ± sem). These values are in good agreement with measurements made by single particle tracking for GluR2 freely diffusing in the plasma membrane of neurons (Borgdorff and Choquet, 2002). The two other populations correspond to molecules which were either immobile (8 %, $D<7\times10^{-3}$ µm$^2$/s, n=25), or diffused slowly (26 % with D lower than 0.1 µm$^2$/s, mean 0.05±0.01 µm$^2$/s, n= 82, see inset of Fig. 2C). In stained synaptic sites, receptors could be separated into two populations (Fig. 2D). Half of the receptors were immobile, whereas the second half diffused with a diffusion coefficient between $1.5\times10^{-2}$ and 0.1 µm$^2$/s (mean 0.054±0.005 µm$^2$/s, n=100). There are thus several populations of GluR2 in terms of diffusion characteristics (from free Brownian diffusion to diffusion in domains), whose proportions vary greatly with respect to synaptic location.

At a given location (synaptic or extra-synaptic), AMPARs with different mobilities were found. In particular, both mobile and immobile receptors were observed successively at the same synaptic site within the same recording sequence in 21 cases, showing that the two receptor's behaviors do not arise from receptors present in separate types of synapses. Moreover, these observations show that on the time scale of our experiments, the movement of synaptic AMPARs is not that of the whole PSDs.

In order to analytically characterize the diffusion properties of each sub-population at each location, we used a second approach based on the distribution of the squared displacements of the molecules (Schuetz et al., 1997). This approach allows us to unravel and analyze multiple diffusion types in each compartment without having to classify the MSDs of the individual molecules, thus avoiding possible bias by an arbitrary sorting. As a result, three categories of receptor movements were also found by this analytical analysis (see Fig. 2S in supplementary material), characterized by the time dependence of their *mean* MSD, $\langle r_i^2(\tau)\rangle$, i=0 to 2. Fast mobile receptors (i=2) were exclusively found in extra-synaptic regions, while the slowly mobile (i=1) and immobile (i=0) receptors were mainly found in synaptic domains (Fig. 2E-F and supplementary material). Thus, two independent analysis protocols, i.e. distributions of individual D values (Fig. 2C-D) and distributions of squared displacements (Fig. 2S in supplementary material and Fig. 2E-F), establish the existence of different receptor populations in terms of mobility. Moreover, the latter analysis paradigm establishes that not only the mobilities, but also the type of diffusion are strikingly different for the two mobile populations. On the one hand $\langle r_2^2(\tau)\rangle$ is linear with time, indicating that these extra-synaptic receptors undergo free Brownian diffusion (at least up to 330 ms, Fig.

2E) with a mean diffusion constant of $0.37 \pm 0.04 \mu m^2/s$. On the other hand, $\langle r_1^2(\tau) \rangle$ saturates with time (Fig. 2F), a signature of spatially restricted diffusion (Kusumi et al., 1993; Simson et al., 1998). These were found, with very similar properties, both in the synaptic and extra-synaptic receptor populations, although in different proportions (45 ± 5% for synaptic and 25 ± 5% for extra-synaptic receptors, not shown). The diffusion coefficient $D_1$ and the diameter L of the domain within which diffusion is restricted can be derived (Kusumi et al., 1993) from :

$$\langle r_1^2(\tau) \rangle = \frac{L^2}{3}\left(1 - \exp\left(\frac{-12 D_1 \tau}{L^2}\right)\right) \quad (1)$$

Least square fitting of the data by equation (1) gives $D_1 = 6 \pm 2 \times 10^{-2}$ $\mu m^2/s$ and a domain size $L = 300 \pm 20$ nm. The domain size is in good quantitative agreement with the synaptic sizes given by electron microscopy (Schikorski and Stevens, 1997; Takumi et al., 1999), suggesting that mobile receptors explore the whole postsynaptic domain. Receptors that we labeled as "extra-synaptic" and which nevertheless displayed restricted diffusion could in fact pertain to unstained synapses. Alternatively, we have previously shown that extra-synaptic receptors aggregated by scaffolding proteins display a similarly restricted mobility (Meier et al., 2001; Sergé et al., 2002).

**Contribution of endocytosis to receptor mobility**

GluR2 containing AMPARs are known to undergo continuous endo/exocytosis and recycling (Carroll et al., 2001). Our experiments did not detect newly exocytosed receptors as no free antibody was present during the recordings. By contrast, antibodies are known to be endocytosed together with the receptors to which they are bound (e.g.(Luscher et al., 1999)). We first measured by imunohistochemistry on live cells the global level of endocytosis of antibody-tagged receptors. Neurons were incubated 10 minutes with anti-GluR2 at 37 °C, washed, and further incubated 15 minutes at 37°C before being fixed and stained for surface and endocytosed receptors (see methods). We found that 24 ± 5 % (mean ± s.e.m., 11 neurons) of the receptors were endocytosed in neurites. Then, to investigate specifically the mobility of endocytosed receptors, we first incubated labeled cells for 30 minutes at 37°C to allow endocytosis to occur, then removed surface labelling by acid wash prior to performing single molecule experiments. Surface staining was decreased by 80 ± 9 % (n=10 neurons) by this treatment. The mobility of the internalised receptors is shown Fig. 3. We found that after acid wash, 79 % of the total receptors (synaptic plus extra synaptic) are immobile, as compared to 25 % in control recordings.

To further investigate the contribution of receptor endocytosis to the proportion of immobile receptors, we performed temperature block of endo/exocytosis in the presence of TTX. First, in immunohistochemistry experiments, we found that the percentage of endocytosis after 15 minutes at 20 °C dropped to 10 ± 4 % (8 neurons). Second, 433 single-molecule trajectories from 20 different live neurons were recorded at 37°C after antibody incubation at 20°C to reduce the amount of internalized receptors and maintain the same recording conditions. We also analyzed 422 trajectories from 20 different neurons with both antibody incubation and recordings at 20°C. The mean diffusion constants of the mobile population of receptors in synaptic sites was not different between 37°C and 20°C (Fig. 3C). However it decreased by a factor of 3 at 20 °C in the extra-synaptic regions (including all mobile extra-synaptic receptors it varied from 0.32±0.04 µm$^2$/s, n=207 at 37°C to 0.11±0.04 µm$^2$/s, n=82 at 20°C). This shows that diffusion of receptors at synaptic sites is not limited by the viscosity of the membrane, as it is likely to be in extra-synaptic regions. This decrease in diffusion constants in extra-synaptic regions led to an apparent 2 fold increase in the proportion of receptors counted as immobile (D<7.10-3µm2/s, not shown). By contrast, as the mean diffusion of receptors in the synaptic regions did not change with temperature, we could directly compare the evolution in the proportion of immobile receptors. It decreased by over 40 % when going from 37°C to 20°C (see Fig. 3D). This decrease is larger than what could be expected from a simple block of endocytosis and may arise from additional phenomenon. In any case, this confirms that endocytosed receptors belong to the immobile population. All data presented below were obtained from recordings performed at 37°C after antibody incubations at 20°C.

### Regulation of AMPARs mobility inside synapses

Postsynaptic plasticity of glutamatergic synapses is mediated in large part by the regulation of AMPARs trafficking (reviewed in (Barry and Ziff, 2002; Braithwaite et al., 2000; Carroll et al., 2001; Malinow and Malenka, 2002; Sheng and Kim, 2002)). Protocols that induce plasticity of synaptic transmission in culture result in changes of AMPARs concentration at synapses and are thought to mimic at the molecular level the processes of LTP and LTD (Beattie et al., 2000; Carroll et al., 1999a; Lin et al., 2000; Lu et al., 2001; Passafaro et al., 2001). Changes in AMPARs numbers at synapses have been mainly attributed to changes in endocytosis or exocytosis of receptors. These membrane traffic events are likely to occur outside PSDs (Blanpied et al., 2002; Passafaro et al., 2001), which implies that receptor diffusion in the plane of the plasma membrane should participate to the changes in synaptic receptor numbers. Here, we analyze whether AMPARs diffusion inside and outside synapses is regulated by glutamate application and changes in synaptic activity.

We used bath application of glutamate to decrease the number of surface expressed AMPARs (Beattie et al., 2000; Carroll et al., 1999a) through mechanisms that may be shared by LTD (Carroll et al., 1999b; Man et al., 2000; Wang and Linden, 2000). This protocol is referred herein as "Glut". Conversely, we also used activation of synaptic release of glutamate through blocking inhibitory neurotransmission by biccuculine and strychnine, together with potentiation of NMDARs by glycine to increase the number of surface AMPARs through mechanisms that may be similar to those of LTP (Lu et al., 2001; Passafaro et al., 2001). This protocol is referred herein as "Bic/Gly". Our control condition was, as previously, recordings in the presence of TTX. As a further resting condition, we also used intracellular BAPTA to chelate pre and postsynaptic intracellular calcium.

We first verified by immunohistochemistry on live neurons (data not shown) that these protocols induced reciprocal changes in surface expression of AMPARs (Beattie et al., 2000; Carroll et al., 1999a; Lin et al., 2000; Lu et al., 2001; Passafaro et al., 2001). Indeed, bath application of 100 µM glutamate induced a 85% increase in the percentage of endocytosed AMPARs within 15 minutes. This corresponded to a loss of 22 % of the surface receptors. Bath application for 5 minutes of 20 µM biccuculine and 1 µM strychnine together with 200 µM glycine induced a 59 % increase  (S.D. = 19 %, n=10) in surface AMPARs. We then studied the effect of these protocols on receptor mobility at the single molecule level. In a first series of experiments, we first labeled surface receptors at 20°C for 10 minutes and then applied the different pharmacological agents during the recordings  (Fig. 4A). The diffusion of extra-synaptic receptors was not significantly modified by the different treatments  (Fig. 4B). In contrast, bath glutamate induced a strong  (55 %) increase of the mean diffusion constant of AMPARs inside synapses (Fig. 4C). This was accompanied by a 30 % reduction in the proportion of immobile synaptic receptors (Fig. 4D). This is surprising since glutamate promotes AMPARs endocytosis and endocytosed receptors are mostly immobile (Fig. 3B). A temptative explanation for this discrepancy is that glutamate treatment induces the accumulation of endocytotic vesicles mainly in the cell body rather than in neuritis, as previously published (data not shown and Beattie et al., 2000). Intracellular BAPTA decreased by 30 % the percentage of immobile synaptic receptors, in accordance with the BAPTA induced decrease in AMPARs stabilization previously reported (Borgdorff and Choquet, 2002). We found no effect on either parameters of the of Bic/Gly protocol when compared to the control TTX condition.

The absence of effect of the Bic/Gly treatment can be surprising given the efficiency of this protocol in increasing the number of surface AMPARs. However, in the condition shown Fig. 4, receptor labeling was performed before Bic/Gly treatment, thus newly inserted receptors were not labeled. Furthermore, previous studies have indicated that upon this type of treatment, receptors might be first exocytosed outside synapses and accumulated at synapses with a delay (Passafaro et al., 2001). We thus delayed the labeling of the receptors to 5 and 40 minutes after application of Bic/Gly and then studied their mobility (Fig. 5). The diffusion coefficients of synaptic or extrasynaptic receptors did not vary much in these different conditions (Fig. 5B et 5C). In contrast, we found a strong transient reduction after 5 minutes of Bic/Gly treatment in the proportion of immobile synaptic receptors (Fig. 5D). Newly inserted receptors are thus initially diffusive and then stabilized at synaptic sites.

**Regulation of receptor localization near synapses**

During glutamate application, we were surprised to qualitatively observe a large number of highly mobile molecules right next to synapses. We thus quantified the localization of receptors in the various conditions and compared their distributions. The distribution *S (r)* of receptors in the presence of bath applied glutamate shown on Fig. 6A, was compared (Fig. 6B) to the control one previously shown on Fig. 1G. In the presence of glutamate, we indeed found a strong increase in the proportion of receptors present in the periphery of synapses, in an annulus about 400 to 800 nm from the center of the synapses. The mobility of these juxtasynaptic AMPARs (D=0.24 ± 0.02 $\mu m^2/s$) is however not different from that of the other extra-synaptic receptors (D=0.25 ± 0.02

µm$^2$/s). We compared the mean percentage of receptors present in this juxtasynaptic annulus in the different conditions studied above (Fig. 6D). In the presence of glutamate, the proportion of juxtasynaptic AMPARs was twice that observed under TTX. No significant change in this proportion was observed in the presence of BAPTA or for 0 or 5 minutes of Bic/Gly treatment. Strikingly, we found a 40 % reduction decrease in the proportion of juxtasynaptic receptors after 40 minutes of Bic/Gly treatment.

**DISCUSSION**

Using single molecule imaging methods, we could image the movement of native AMPARs in synapses. The different behaviors of receptors movements both inside and outside synapses, as well as the topological distribution of the receptors in live neurons could be sorted out by studying over 5000 single molecule trajectories. We found that about half of synaptic receptors are mobile and that both the proportion and the amount of mobility are regulated by pharmacological treatments that modify receptor accumulation at synapses.

Individual fluorophores bear unique signatures, such as one step photobleaching, well defined signals and diffraction limited spots (Weiss, 1999). These allow us to ascertain that we are tracking single fluorescently tagged anti-GluR2 antibodies. Single molecule imaging can report receptor localization and movement. First, the good signal to noise ratio allow for a 45 nm pointing accuracy (Schmidt et al., 1995; Thompson et al., 2002). Second, in the extra-synaptic membrane, the presence of the antibody on the AMPARs will not *per se* modify receptor movements, as viscosity of the membrane and cytoplasm will by far dominate to limit diffusion. Third, the antibody itself did not promote aggregation of the receptors. However, the antibodies may report the movement of either individual or native clusters of receptors. We have no means to determine how many receptors would compose such native microaggregates, as both theoretical and experimental data have supported the notion that diffusion in a two dimensional space is barely affected by the size of the diffusing object. This is exemplified by the fact that we observed comparable rates of free diffusion by single molecule imaging and tracking single receptor-bound 500 nm particles (Borgdorff and Choquet, 2002).

To identify synapses on live cells, we used FM1-43 to stain recycling synaptic vesicles and rhodamin 123 to stain mitochondrias, which are enriched in presynaptic terminals. These approaches correlate very well as both markers colocalized at around 80% in double staining experiments (data not shown). The density distribution of receptors, as seen by single molecule imaging, peaked at about 100 nm from the synapse center and decreased with a shallow slope toward the extra-synaptic membrane. Although single molecules are pointed with a 45 nm accuracy, presynaptic terminals are localized with less precision due to a bulkier labeling, even if FM1-43 has been reported as a precise marker of presynaptic terminals (Murthy et al., 2001). It would thus be hazardous to directly compare our distributions with electron microscopy data. However, it is interesting to note that the area of receptor enrichment that we observe is roughly on the same range, although on the high end, as that observed with electron microscopy (Schikorski and Stevens, 1997; Takumi et al., 1999).

Antibody binding to AMPARs could potentially affect receptor movement in the synaptic cleft because of steric hindrance by extracellular molecules. This potential effect is difficult to appreciate. Indeed, on the one hand the molecule density and arrangement in the synaptic cleft is not known. On the other hand, the size of the extracellular domain of AMPARs (Armstrong and Gouaux, 2000) is by itself rather large (about 10 by 5 nm for one domain) and is comparable to that of an IgG molecule. In any case, our approach gives an upper limit on AMPARs mobility inside synapses.

Most receptors displayed Brownian movements outside synapses, as expected from freely diffusing receptors and previous studies (Borgdorff and Choquet, 2002). Mobile receptors inside synapses displayed a confined movement. This is a further indirect argument that molecules colocalized with presynaptic stainings are indeed synaptic receptors. In synapses, receptor movements are confined to domains of mean radius 150 nm. This in good agreement with the reported size of PSDs as seen by EM (Schikorski and Stevens, 1997; Takumi et al., 1999). The instantaneous diffusion of receptors in the domains was on average 0.6 $\mu m^2/s$, five fold lower than in the extrasynaptic membrane and 10 times higher than our experimental threshold for detection of mobile receptors. The reduction of receptor mobility and the process of confinement in synapses could have similar or different origins. Both are likely to arise from interactions with molecules present at high density in the PSDs (Scannevin and Huganir, 2000). These interactions could be either specific through short unresolved transient binding to associated scaffold molecules or unspecific and leading to frictional slowing of the receptors. This is compatible with the observation that lowering temperature does not affect synaptic receptor diffusion. Confinement of receptors could also arise from corralling by barriers present at the periphery of the PSD. These barriers could be for example adhesion proteins such as cadherins which are enriched in an annulus around PSDs (Uchida et al., 1996). These barriers would have to be permeable as we observed receptors entering and leaving synapses through lateral diffusion. In synapses, we also observed a large proportion of immobile receptors that are most likely bound to scaffold proteins acting as slots. Altogether, a unified picture of the postsynaptic density could be one where receptors are immobilized for transient periods of time related to the receptor-scaffold affinity. When unbound, receptors could diffuse in the space between slots at a relatively high rate and easily escape or reenter the postsynaptic space. Trafficking of receptors in and out of synapses through lateral diffusion is likely to be a general phenomenon; NMDA receptors have been shown by electrophysiological recordings to exchange between synaptic and extra-synaptic sites through traffic in the plasma membrane (Tovar and Westbrook, 2002).

In single particle tracking studies, we had previously observed that receptor-bound particles stop reversibly at or near synapses (Borgdorff and Choquet, 2002). In those studies the particle size (500 nm) prevented a more precise determination of the localization of immobilization sites. The present study indicates that receptors in the juxtasynaptic region have diffusion properties indistinguishable from those of other extra-synaptic receptors. Thus the majority of sites of receptor immobilization are just in front of presynaptic terminals.

A number of protocols related to synaptic plasticity modify the accumulation of receptors at synapses. These modifications have mainly been viewed in the frame of regulation of endocytic and exocytic mechanisms (Barry and Ziff, 2002; Carroll et al., 2001). Here, we found that bath application of glutamate, which induces rapid depletion of AMPARs from synapses (Carroll et al., 1999b; Lissin et al., 1999), increases synaptic receptor diffusion rate, decreases the proportion of immobile synaptic receptors and increases the proportion of receptors in an area surrounding synapses. These results strongly suggest that glutamate-induced loss of postsynaptic receptors is due primarily to their escape from the PSDs through lateral diffusion. This would then be followed by their endocytosis in the extrasynaptic membrane. We have shown that endocytosed receptors are mainly immobile. That glutamate induced endocytosed receptors are mainly concentrated in cell bodies (Beattie et al., 2000) may explain why they escape our observation in neurites. Escape of AMPARs from the PSD could involve either the disruption of the post synaptic actin cytoskeleton (Zhou et al., 2001) or a specific decrease in receptor affinity for a scaffold element. This would explain both the increased diffusion coefficient, by lowering the rate of fast transient binding of receptors to scaffold elements, and the decreased proportion of immobile receptors. The increase in the proportion of juxtasynaptic receptors in the presence of glutamate may correspond to receptors flowing out of the PSDs.

Inhibition of inhibitory neurotransmission together with potentiation of NMDARs activity, which results in increased surface expression of AMPARs (Lu et al., 2001; Passafaro et al., 2001), exhibit time-dependent effects on receptor mobility and localization. In the first minutes, we found that it mainly decreases the proportion of immobile synaptic receptors. After 40 minutes, both diffusion rates and percentage of immobile synaptic receptors are back to control values and the proportion of juxtasynaptic receptors is decreased. This observation relates to previous results on the fate of newly exocytosed AMPARs; using cleavable extracellular tags, it was observed that at early times after exocytosis, new GluR1 containing AMPARs are diffusively distributed along dendrites. This is followed by their lateral translocation and accumulation into synapses (Passafaro et al., 2001). GluR2 subunits seemed to be addressed directly at synapses. In our experiments, we are following the movement of native GluR2 containing AMPARs whose precise subunit composition is unknown. They could also contain GluR1, GluR3 or GluR4, which could impose different trafficking behaviors (Barry and Ziff, 2002; Malinow and Malenka, 2002). In any case, our data suggest that at the level of synapses themselves, newly added receptors are initially diffusive and then stabilized over time.

We found that the proportion of juxtasynaptic receptors varies inversely with the degree of receptor stabilization at the synapse. This supports the notion that the juxtasynaptic region around synapses represents not only a transit zone for receptors entering and leaving synapses through lateral diffusion, but also a reserve pool zone where receptors are available for recruitment at synapses. The AMPAR–associated protein stargazin might be a molecular determinant of this regulation of AMPAR trafficking between the synaptic and extrasynaptic membrane through trapping of diffusing AMPARs by binding to synaptic PSD-95 (Schnell et al., 2002). In this model, the pools of extra-synaptic and synaptic receptors are in a dynamic equilibrium. The degree of accumulation at the synapse would be set both by the total surface number of receptors and the residency time of the receptors in the synapse. It will be important to determine these times which relate to the affinity of the receptors for the scaffold.

## MATERIALS AND METHODS

### Microscopy and single-molecule detection

A custom wide field single-molecule fluorescence inverted microscope equipped with a 100x oil-immersion objective (NA=1.4) was used (IX70 Olympus, Bordeaux, France). The samples were illuminated for 30 ms at a wavelength of 633 nm by an He-Ne laser (JDS Uniphase, Manteca, CA, USA) at a rate of 33 Hz. Use of a defocusing lens permitted to illuminate a surface of 20x20 $\mu m^2$ with a mean illuminating intensity of 7±1 $kW/cm^2$. An appropriate filter combination (DCLP650, HQ575/50, Chroma Technology, Brattleboro, USA) permitted the detection of individual fluorophore by an intensified CCD camera system (Pentamax Princeton Instruments, Trenton, NY, USA). The total detection efficiency of the experimental setup was ~5%. Using the same excitation path with another filter combination (DCLP498, Chroma Technology, and BA515, Omega Optical, Brattleboro, USA), FM1-43 and rhodamine 123 was excited with the 488 nm line of an $Ar^+$ laser (Spectra Physics, Les Ulis, France) at an illuminating intensity of 2 $kW/cm^2$. The images were then recorded on the same CCD camera at 33Hz.

### Cell culture, GluR2 and synapses staining

Hippocampal neurons from 18 days old rat embryos were cultured on glass coverslips following the Banker technique as previously (Borgdorff and Choquet, 2002). For staining in TTX condition, neurons were incubated at 37°C (or 20°C) for 30 seconds with 5 µM FM1-43 (Molecular Probes, Eugene, OR) in a culture medium supplemented with 40 mM KCl. They were then rinsed 1 minute in culture medium supplemented with 10 mM Hepes and 1µM TTX (recording medium) before being incubated 10 min at 37°C (or 20°C) with 10 µg/ml anti-GluR2-Cy5 or anti-GluR2-Alexa647 (see supplementary). In control experiments, we verified that most FM1-43 staining disappeared upon further depolarization by 40 mM KCl (not shown). In the two other conditions (see below), neurons were first incubated at 37°C for 5 min with 2 µM rhodamine123. They were then fast rinsed and incubated at room temperature for 10 min with the labeled antibodies. After fast rinses, the coverslips were mounted in a custom chamber with culture medium supplemented with 20 mM Hepes. The medium also contained either 1 µM TTX, 100 µM glutamate or a combination of 20 µM Biccuculine, 1 µM strychnine and 200 µM glycine. For the BAPTA condition, neurons were first incubated with 5 µM BAPTA-AM at 37°C for 10 min before the antibody labeling, then rinsed at 37°C during 2min and recorded in the presence of 1 µM TTX. All data were taken within 20 minutes after the last rinse. On a few occasions, we observed rapid (1µm/s) movements of FM1-43 stains that we attributed to vesicle trafficking and were therefore discarded from our analysis. Rhodamine 123 concentrates in regions with high mitochondrial density such as presynaptic terminals. We checked that the FM1-43 and rhodamine 123 staining colocalized well (not shown). Specificity of the anti-GluR2 labeling was confirmed as virtually no staining was observed with the anti-GluR2 on several cell lines not expressing GluR2. We also verified by trypan blue exclusion that 15 minutes incubation with glutamate did not lead to immediate cell death (cell viability in control and glutamate conditions $74 \pm 10$ %, n=20 and $72 \pm 14$ %, n=16 fields, respectively, n=355 and 209 neurons). For acid wash to remove surface labeling, live neurons were first labeled as above with anti-GluR2, incubated 30 minutes at 37°C in culture medium and then washed 2 minutes at 4°C in culture medium at pH 2 just before the experiment. Neurons were still metabolically active after this treatment as for example we could image normal intracellular transport of mitochondrias (data not shown).

For immunocytochemistry, neurons were incubated at 37°C or 20°C, as indicated in the text or figures, with anti-GluR2 for 10 minutes. Neurons were then either fixed immediately with 4%PFA and sucrose or rinsed, maintained live further for 15 minutes and then fixed and processed with secondary antibodies. For detection of surface and endocytosed receptors, fixed cells were first incubated 45 minutes with 10 µg/ml secondary alexa-568 (molecular probes, Leiden, The Netherlands) anti-mouse antibody to saturate all surface bound anti-GluR2s. Saturation was verified in control experiments (not shown). Cells were then permeabilized and incubated 45 minutes with 10 µg/ml secondary alexa-488 anti-mouse antibody to reveal endocytosed receptors. Green and red flurorescent images were quantified using metamorph (Universal Imaging, Downingtown PA). The comparison of the levels of receptor clustering in live and fixed cells was measured as in (Sergé et al., 2002).

**Trajectory construction**

The spatial distribution of the signals on the CCD originating from individual molecules was fitted to a two-dimensional Gaussian surface with a full-width at half-maximum of $360 \pm 40$ nm, given by the point-spread function of our apparatus. The two-dimensional trajectories of single molecules in the plane of focus were constructed by correlation analysis between consecutive images using a Vogel algorithm (Schuetz et al., 1997). Only trajectories containing at least 3 points were retained. This sets the minimum diffusion constant detectable with our setup, $7 \times 10^{-3}$ µm$^2$/s, which corresponds to the binning on the histograms. A systematic subtraction of 0.002 µm$^2$ is used to reject the bias induced by the pointing accuracy of the MSD.

**Synaptic localization of AMPARs**

The synaptic staining maxima (FM1-43 or rhod 123) were determined by fitting the fluorescence spots corresponding to presynaptic terminal with two dimensional Gaussians. The error on the fit was in the order of 60 nm, depending on the asymmetry of synaptic shape. Thus, the distance from synapse center to a molecule was measured with a precision of about 80 nm, given the 45 nm pointing accuracy on molecules. The surfacic proportion, *S (r)*, is defined as the number of molecules detected between distances, *r*, and, *r+dr*, from the center of the nearest synapse divided by the total number of molecules and by the elementary surface of width *dr*. The results are shown on Fig. 1G and 6D for steps *dr=40nm* and a mean neurite width of 1µm (in a simple approximation, the elementary surface to take into account is equal to *2πrdr* for *r*<500nm and constant for *r*>500nm).

## ACKNOWLEDGEMENTS

We wish to thank Philippe Ascher, Antoine Triller, Christophe Mulle, and Thomas Schmidt for their precious comments on this manuscript, Michel Orrit for initiating this collaborative work and Laurent Groc for suggesting the acid wash experiment. We thank Dr. El Mestikawy for the gift of the anti BNPI antibody. This work was supported by grants from the CNRS, the Conseil Régional d'Aquitaine and the ministère de la recherche.

The authors declare that they have no competing interests.

Anderson, C.M., Georgiou, G.N., Morrison, I.E., Stevenson, G.V. and Cherry, R.J. (1992) Tracking of cell surface receptors by fluorescence digital imaging microscopy using a charge-coupled device camera. Low-density lipoprotein and influenza virus receptor mobility at 4 degrees C. *J Cell Sci*, **101**, 415-425.

Armstrong, N. and Gouaux, E. (2000) Mechanisms for activation and antagonism of an AMPA-sensitive glutamate receptor: crystal structures of the GluR2 ligand binding core. *Neuron*, **28**, 165-181.

Barry, M.F. and Ziff, E.B. (2002) Receptor trafficking and the plasticity of excitatory synapses. *Curr Opin Neurobiol*, **12**, 279-286.

Beattie, E.C., Carroll, R.C., Yu, X., Morishita, W., Yasuda, H., von Zastrow, M. and Malenka, R.C. (2000) Regulation of AMPA receptor endocytosis by a signaling mechanism shared with LTD. *Nat Neurosci*, **3**, 1291-1300.

Blanpied, T.A., Scott, D.B. and Ehlers, M.D. (2002) Dynamics and regulation of clathrin coats at specialized endocytic zones of dendrites and spines. *Neuron*, **36**, 435-449.

Borgdorff, A. and Choquet, D. (2002) Regulation of AMPA receptor lateral movement. *Nature*, **417**, 649–653.

Braithwaite, S.P., Meyer, G. and Henley, J.M. (2000) Interactions between AMPA receptors and intracellular proteins. *Neuropharmacology*, **39**, 919-930.

Carroll, R.C., Beattie, E.C., von Zastrow, M. and Malenka, R.C. (2001) Role of ampa receptor endocytosis in synaptic plasticity. *Nat Rev Neurosci*, **2**, 315-324.

Carroll, R.C., Beattie, E.C., Xia, H., Luscher, C., Altschuler, Y., Nicoll, R.A., Malenka, R.C. and von Zastrow, M. (1999a) Dynamin-dependent endocytosis of ionotropic glutamate receptors. *Proc Natl Acad Sci U S A*, **96**, 14112-14117.

Carroll, R.C., Lissin, D.V., von Zastrow, M., Nicoll, R.A. and Malenka, R.C. (1999b) Rapid redistribution of glutamate receptors contributes to long-term depression in hippocampal cultures. *Nat Neurosci*, **2**, 454-460.

Chen, L., Chetkovich, D.M., Petralia, R.S., Sweeney, N.T., Kawasaki, Y., Wenthold, R.J., Bredt, D.S. and Nicoll, R.A. (2000) Stargazing regulates synaptic targeting of AMPA receptors by two distinct mechanisms. *Nature*, **408**, 936-943.

Dickson, R.M., Norris, D.J., Tzeng, Y.L. and Moerner, W.E. (1996) Three-dimensional imaging of single molecules solvated in pores of poly(acrylamide) gels. *Science*, **274**, 966-969.

Dingledine, R., Borges, K., Bowie, D. and Traynelis, S.F. (1999) The glutamate receptor ion channels. *Pharmacol Rev*, **51**, 7-61.

Ehlers, M.D. (2000) Reinsertion or degradation of AMPA receptors determined by activity- dependent endocytic sorting. *Neuron*, **28**, 511-525.

Kusumi, A., Sako, Y. and Yamamoto, M. (1993) Confined lateral diffusion of membrane receptors as studied by single particle tracking (nanovid microscopy). Effects of calcium-induced differentiation in cultured epithelial cells. *Biophys J*, **65**, 2021-2040.

Lin, J.W., Ju, W., Foster, K., Lee, S.H., Ahmadian, G., Wyszynski, M., Wang, Y.T. and Sheng, M. (2000) Distinct molecular mechanisms and divergent endocytotic pathways of AMPA receptor internalization. *Nat Neurosci*, **3**, 1282-1290.

Lissin, D.V., Carroll, R.C., Nicoll, R.A., Malenka, R.C. and von Zastrow, M. (1999) Rapid, activation-induced redistribution of ionotropic glutamate receptors in cultured hippocampal neurons. *J Neurosci*, **19**, 1263-1272.

Lu, W., Man, H., Ju, W., Trimble, W.S., MacDonald, J.F. and Wang, Y.T. (2001) Activation of synaptic NMDA receptors induces membrane insertion of new AMPA receptors and LTP in cultured hippocampal neurons. *Neuron*, **29**, 243-254.

Luscher, C., Xia, H., Beattie, E.C., Carroll, R.C., von Zastrow, M., Malenka, R.C. and Nicoll, R.A. (1999) Role of AMPA receptor cycling in synaptic transmission and plasticity. *Neuron*, **24**, 649-658.

Malinow, R. and Malenka, R.C. (2002) AMPA receptor trafficking and synaptic plasticity. *Annu Rev Neurosci*, **25**, 103-126.

Man, Y.H., Lin, J.W., Ju, W.H., Ahmadian, G., Liu, L., Becker, L.E., Sheng, M. and Wang. Y, T. (2000) Regulation of AMPA receptor-mediated synaptic transmission by clathrin-dependent receptor internalization. *Neuron*, **25**, 649-662.

Meier, J., Vannier, C., Sergé, A., Triller, A. and Choquet, D. (2001) Fast and reversible trapping of surface glycine receptors by gephyrin. *Nature Neuroscience*, **4**, 253-260.

Murthy, V.N., Schikorski, T., Stevens, C.F. and Zhu, Y. (2001) Inactivity produces increases in neurotransmitter release and synapse size. *Neuron*, **32**, 673-682.

Noel, J., Ralph, G.S., Pickard, L., Williams, J., Molnar, E., Uney, J.B., Collingridge, G.L. and Henley, J.M. (1999) Surface expression of AMPA receptors in hippocampal neurons is regulated by an NSF-dependent mechanism. *Neuron*, **23**, 365-376.

Nusser, Z. (2000) AMPA and NMDA receptors: similarities and differences in their synaptic distribution. *Curr Opin Neurobiol*, **10**, 337-341.

Passafaro, M., Piech, V. and Sheng, M. (2001) Subunit-specific temporal and spatial patterns of AMPA receptor exocytosis in hippocampal neurons. *Nat Neurosci*, **4**, 917-926.

Scannevin, R.H. and Huganir, R.L. (2000) Postsynaptic organization and regulation of excitatory synapses. *Nat Rev Neurosci*, **1**, 133-141.

Schikorski, T. and Stevens, C.F. (1997) Quantitative ultrastructural analysis of hippocampal excitatory synapses. *J Neurosci*, **17**, 5858-5867.

Schmidt, T., Schuetz, G.J., Baumgartner, W., Gruber, H.J. and Schindler, H. (1995) Characterization of photophysics and mobility of single molecules in a fluid lipid membrane. *J.Phys.Chem.*, **99**, 17662-17668.

Schnell, E., Sizemore, M., Karimzadegan, S., Chen, L., Bredt, D.S. and Nicoll, R.A. (2002) Direct interactions between PSD-95 and stargazin control synaptic AMPA receptor number. *Proc Natl Acad Sci U S A*, **99**, 13902-13907.

Schuetz, G.J., Schindler, H. and Schmidt, T. (1997) Single-molecule microscopy on model membranes reveals anomalous diffusion. *Biophys.J.*, **73(2)**, 1073-1080.

Schutz, G., Kada, G., Pastushenko, V. and Schindler, H. (2000) Properties of lipid microdomains in a muscle cell membrane visualized by single molecule microscopy. *EMBO J*, **19**, 892-901.

Seisenberger, G., Ried, M.U., Endre{beta}, T., Buning, H., Hallek, M. and Brauchle, C. (2001) Real-Time Single-Molecule Imaging of the Infection Pathway of an Adeno-Associated Virus. *Science*, **294**, 1929-1932.

Sergé, A., Fourgeaud, L., Hémar, A. and Choquet, D. (2002) Receptor activation and homer differentially control the lateral mobility of mGluR5 in the neuronal membrane. *J. Neuroscience*, **22**, 3910-3920.

Sheng, M. and Kim, M.J. (2002) Postsynaptic signaling and plasticity mechanisms. *Science*, **298**, 776-780.

Simson, R., Yang, B., Moore, S.E., Doherty, P., Walsh, F.S. and Jacobson, K.A. (1998) Structural mosaicism on the submicron scale in the plasma membrane. *Biophys J*, **74**, 297-308.

Snyder, E.M., Philpot, B.D., Huber, K.M., Dong, X., Fallon, J.R. and Bear, M.F. (2001) Internalization of ionotropic glutamate receptors in response to mGluR activation. *Nat Neurosci*, **4**, 1079-1085.

Special-Issue. (1999) Frontiers in Chemistry: Single Molecules. *Science*, **283**, 1668-1695.

Takumi, Y., Ramirez-Leon, V., Laake, P., Rinvik, E. and Ottersen, O.P. (1999) Different modes of expression of AMPA and NMDA receptors in hippocampal synapses. *Nat Neurosci*, **2**, 618-624.

Thompson, R.E., Larson, D.R. and Webb, W.W. (2002) Precise nanometer localization analysis for individual fluorescent probes. *Biophys J*, **82**, 2775-2783.

Tovar, K.R. and Westbrook, G.L. (2002) Mobile NMDA receptors at hippocampal synapses. *Neuron*, **34**, 255-264.

Uchida, N., Honjo, Y., Johnson, K.R., Wheelock, M.J. and Takeicji, M. (1996) The catenin/cadherin adhesion system is localized in synaptic junctions bordering translitter release zones. *J.Cell Biol.*, **135**, 767-779.

Ueda, M., Sako, Y., Tanaka, T., Devreotes, P. and Yanagida, T. (2001) Single-molecule analysis of chemotactic signaling in Dictyostelium cells. *Science*, **294**, 864-867.

Wang, Y.T. and Linden, D.J. (2000) Expression of cerebellar long-term depression requires postsynaptic clathrin-mediated endocytosis. *Neuron*, **25**, 635-647.

Weiss, S. (1999) Fluorescence spectroscopy of single biomolecules. *Science*, **283**, 1676-1683.

Zhou, Q., Xiao, M. and Nicoll, R.A. (2001) Contribution of cytoskeleton to the internalization of AMPA receptors. *Proc Natl Acad Sci U S A*, **98**, 1261-1266.

**LEGENDS TO FIGURES**

Figure 1:

Single-molecule fluorescence detection of GluR2-containing AMPARs localization. (**A-C**) Simultaneous images of a neurite of a living neuron as seen by Differential Interference Contrast (**A**) and epifluorescence of FM1-43 on a green channel and Cy5 on a red channel (**B, C**). (**B**) Synaptic sites stained by depolarization-induced uptake of FM1-43. (**C**) Diffraction limited spot image of a single Cy5-anti-GluR2 antibody. The molecule is colocalized with one of the three synaptic sites of Fig. 1B (see also the movies in the supplementary materials). The scale bar is 1µm. (**D**) 3D representation (intensity in the vertical axis) of the fluorescence of the molecule in (**B**). Scale bar is counts per 30ms. (**E**) Recording of the fluorescence intensity of a single Cy5-anti-GluR2 molecule over time displaying the characteristic one-step photobleaching. (**F**) Double staining of surface GluR2 on a live neuron (left and red in the merge) and presynaptic terminals with an anti-vesicular glutamate transporter (middle and green in the merge). (**G**) Histogram of S (r), the surface proportion of the detected AMPARs on live neurons as a function of their distance, r, from the closest maximum in synaptic staining intensity. Data from single molecules recorded in the presence of TTX. AMPARs accumulate at and close to synaptic staining maxima.

Figure 2:

(**A-B**) Illustrative examples of AMPARs movements. (**A**) Examples of trajectories of individual molecules. (#1) is a Cy5-anti-GluR2 fixed on a coverslip. The other trajectories correspond to single Cy5-anti-GluR2 bound to AMPARs in living dendrites. The trajectories recorded in synaptic regions (at less than 300 nm from a local maxima of synaptic staining) are indicated in green. The trajectories recorded in extrasynaptic domains are indicated in red. Trajectories (#2) and (#3) stayed within synaptic sites; trajectory (#4) evolved entirely in the extra synaptic membrane; trajectory (#5) started in an extra synaptic region and then entered a synaptic site. (**B**) Plots of the Mean square displacement (MSD) versus time intervals τ for trajectories shown in (**A**) Trajectories #2 and #3 which were both synaptic, were less mobile than trajectory #1 but differed : #2 corresponds to a slowly mobile receptor in a confined area, #3 to an immobile receptor.

(C-F) statistical analysis of AMPARs movements. (C-D) The differential mobility of synaptic and extra-synaptic AMPARs is shown by the differences in instantaneous diffusion constants. The diffusion constants were measured at 37°C in the presence of TTX. **(C)** Histograms of the instantaneous diffusion constant for 306 AMPARs trajectories detected in extra synaptic regions of more than twenty neurons **(D)** Same histograms for 187 AMPARs trajectories detected in synaptic sites of more than twenty neurons. Binning of the major histograms is 0.075 µm$^2$/s. The insets correspond to the same data with a binning of 0.007 µm$^2$/s. (E-F) Single-molecule analysis of AMPARs diffusion: plots of the mean MSD $<r_1^2(\tau)>$ and $<r_2^2(\tau)>$ derived from the analysis of the square displacements for $\tau = n \times 30$ ms (n=1 to 11). **(E)** $<r_2^2(\tau)>$ is linear with time and reveals free diffusion for fast diffusing extra-synaptic AMPARs. **(F)** $<r_1^2(\tau)>$ are undistinguishable for synaptically_located (filled circle) and slowly diffusing extra synaptic (open squares) trajectories and reveals confined movement.

Figure 3:

**(A-B)** Histograms of the cumulative distribution of instantaneous diffusion constant of synaptic and extrasynaptic receptors in control conditions **(A)** and after acid wash to detect specifically endocytosed AMPARs (493 and 975 trajectories, respectively). Binning as in Fig 2C.

(C-D) Temperature dependence of AMPAR's diffusion. **(C)** Mean instantaneous diffusion constant for (freely diffusing) AMPARs in extra synaptic regions (red, hollow) and for diffusing AMPARs (with restricted diffusion) at synaptic sites (green, filled) as a function of $T_I$ and $T_E$. For both graphs, the values $T_I$ and $T_E$ are indicated on the figure **(D)** Fraction of immobile over mobile receptors in synaptic sites for different incubation temperatures during the labeling by anti-GluR2, $T_I$, or during the experiments, $T_E$ (n=187/7, 210/7 and 210/5 trajectories/experiments for respectively, $T_I / T_E$ = 37°C /37°C, 20°C /37°C and 20°C /20°C, p<0.003 that the values for the first and third conditions are different, Student's *t*-test)..

Figure 4:

Regulation of AMPARs mobility in conditions of synaptic plasticity. **(A)** Protocols: synapse labeling (SL) is followed by antibody incubation for 10 minutes at 20°C (Ab) before single molecule experiments at 37°C for 20 minutes in the presence of different pharmacological agents. **(B)** Mean diffusion constants for mobile extra-synaptic AMPARs in the 4 conditions (n = 356/10, 606/16, 627/15 and 578/21 trajectories/experiments for respectively TTX, BAPTA, Glutamate and Bic/Gly applications). No significant difference is detected. **(C)** Mean diffusion constants (± S.D.) for mobile synaptic AMPARs in the 4 conditions. Glutamate induces a 55 % increase in the mobility of the synaptic AMPARs ($p<0.001$ that the Glut value is different from the 3 other, Student's *t*-test; n = 177/10, 60/16, 169/15 and 54/21 trajectories/experiments for respectively TTX, BAPTA, Glutamate and Bic/Gly applications). **(D)** Fraction of immobile over mobile receptors in synaptic sites for the four different conditions showing a slight but significant decrease of the proportion of immobile synaptic AMPARs during glutamate or BAPTA application ($p<0.05$ that the BAPTA and Glutamate value are different from the 2 others, Student's *t*-test).

Figure 5:

Transient regulation of AMPARs mobility in conditions of Biccuculine/Glycine bath application. **(A)** Protocols: the fourth protocol shown on Fig. 4A is modified by delaying the antibody incubation either 5 or 40 minutes after bath application of biccuculine, strychnine and glycine (bic/Gly). **(B)** Mean diffusion constants for mobile extrasynaptic AMPARs in the 3 conditions. **(C)** Mean diffusion constants for mobile synaptic AMPARs in the 3 conditions. **(D)** Fraction of immobile over mobile receptors in synaptic sites for the three conditions showing a transient decrease of the proportion of immobile synaptic AMPARs at 5 minutes after bic/Gly treatment ($p<0.01$ that the Bic/Gly5 is different from the 2 others, Student's *t*-test; n = 54/21, 64/29 and 70/15 trajectories/experiments for respectively Bic/Gly0, Bic/Gly5 and Bic/Gly40).

Figure 6:

Regulation of the topological distribution of extra-synaptic AMPARs in conditions of synaptic plasticity. **(A)** Surface proportion of the detected AMPARs, S (r), as on Fig.

1G, but in the presence of glutamate. **(B)** The difference (%) between Fig. 1G and Fig. 6D reveals a significant increase of AMPARs in the presence of glutamate in an annulus about 400 to 800 nm from the center of the synapses defined latter as the "juxtasynaptic region". **(C)** Example of a juxta-synaptic trajectory and of the different regions defined in the plasma membrane of the neurons with respect to synaptic staining. Scale bar 500 nm **(D)** Ratio of juxtasynaptic over extra-synaptic AMPARs in the six different conditions shown on Fig 4A and 5A. Glutamate-induced LTD doubles the proportion of juxtasynaptic AMPARs ($p<0.001$ that the Glut value is different from the 5 others, Student's *t*-test) and a 40 % decrease in the proportion of juxtasynaptic receptors after 40 minutes of Bic/Gly is found ($p<0.05$ that the Bic/Glyc40 value is different from the 5 others, Student's *t*-test).

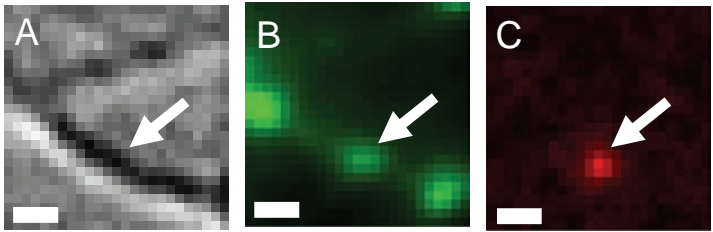
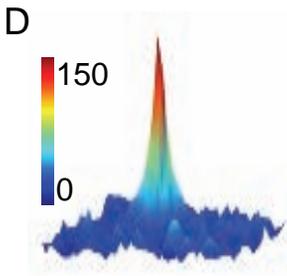
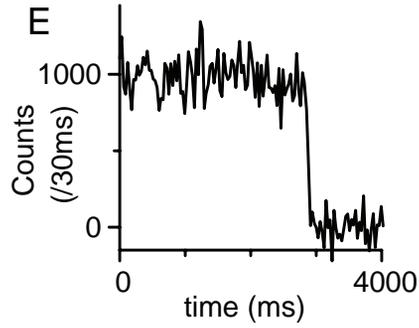
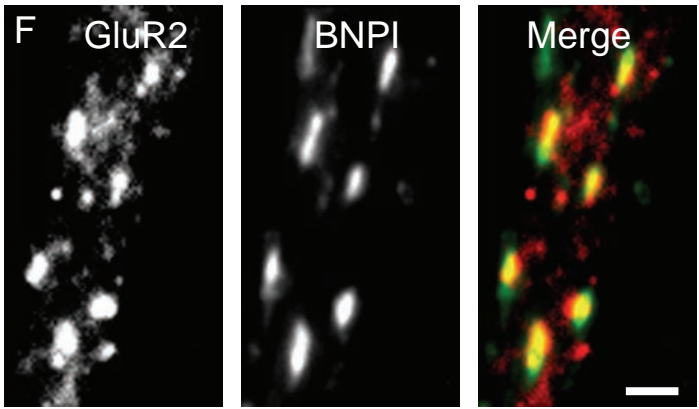
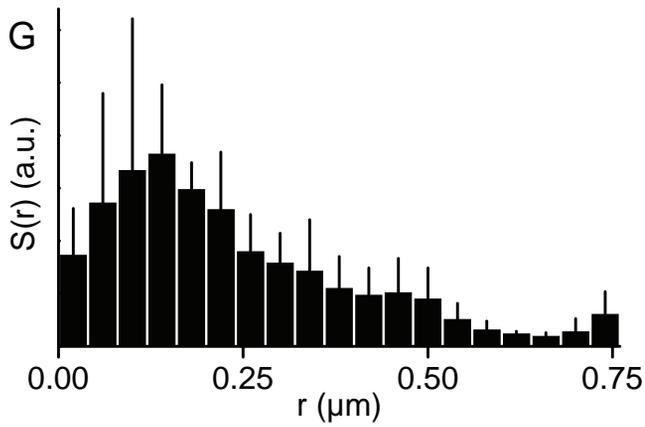

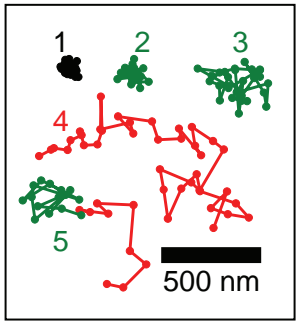
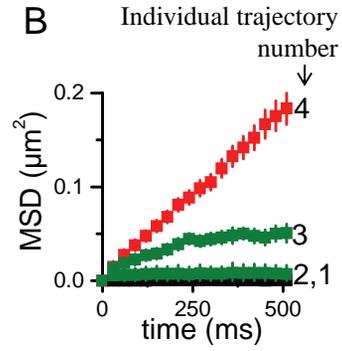
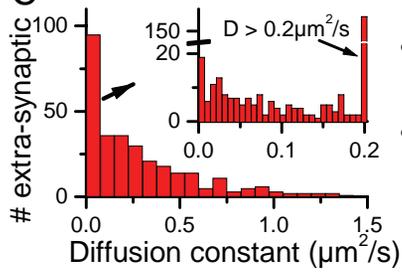
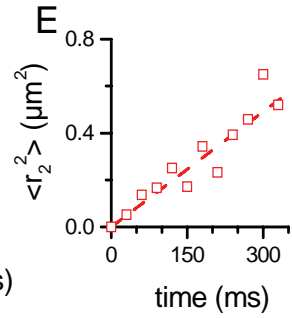
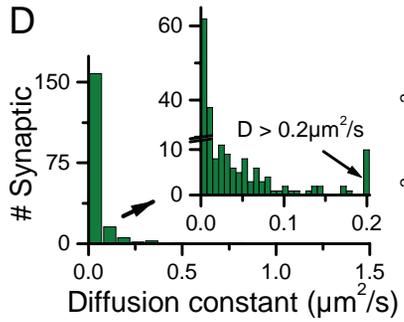
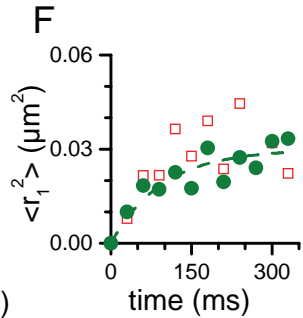

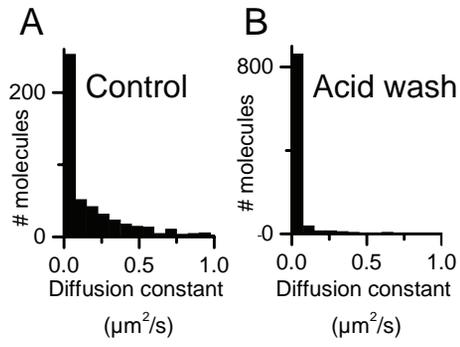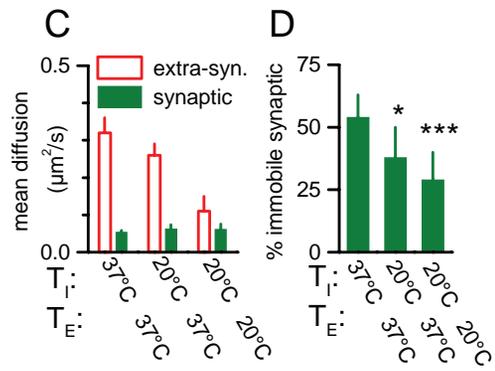

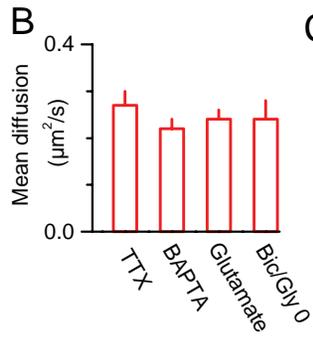
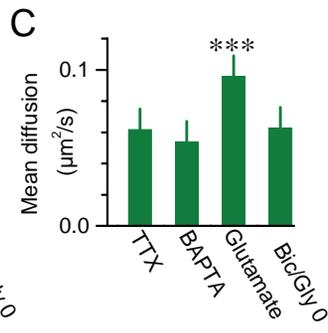
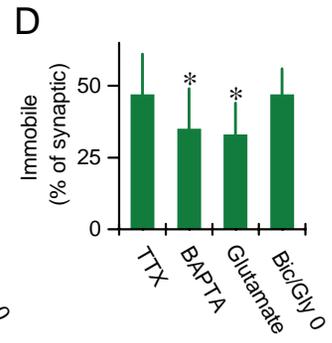

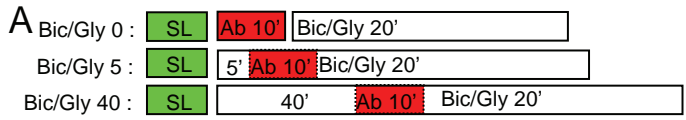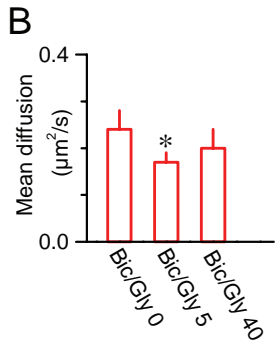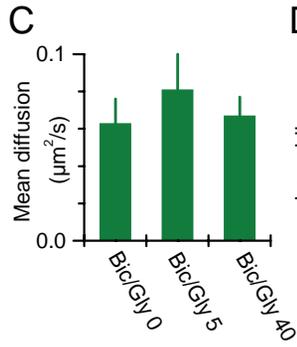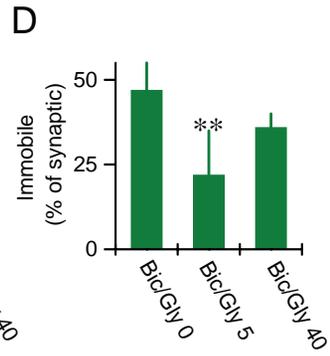

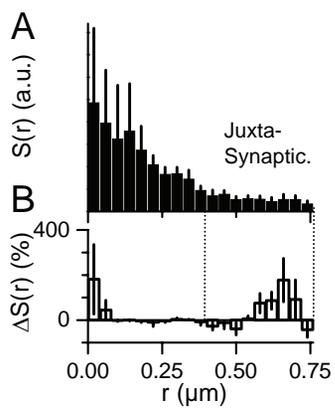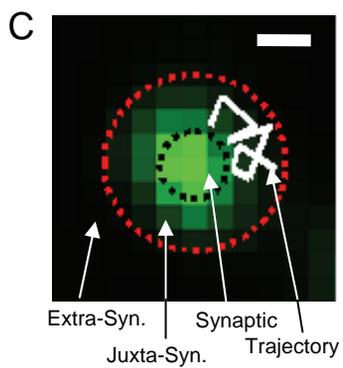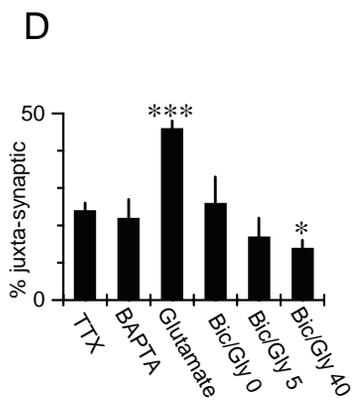